\documentclass[%
prl,
twocolumn,
amsmath,
amssymb,
reprint,
superscriptaddress,
footinbib,
 amsmath,amssymb,
 aps,
]{revtex4-1}

\usepackage{graphicx}
\usepackage{dcolumn}
\usepackage{bm}
\usepackage{natbib}
\usepackage{color}

\begin{document}

\preprint{APS/123-QED}

\title{Nano-Kelvin thermometry and temperature control:  beyond the thermal noise limit}

\author{Wenle Weng}
\email{wlweng@physics.uwa.edu.au}
\affiliation{School of Physics, University of Western Australia, WA 6009, Australia}

\author{James D. Anstie}

\affiliation{Institute for Photonics and Advanced Sensing and School of Chemistry and Physics, University of Adelaide, SA 5005, Australia}
\affiliation{School of Physics, University of Western Australia, WA 6009, Australia}

\author{Thomas M. Stace}
\affiliation{School of Mathematics and Physics, University of Queensland, Brisbane, Queensland 4072, Australia}

\author{Geoff Campbell}
\affiliation{Department of Quantum Science, Australian National University, ACT 0200, Australia}

\author{Fred N. Baynes}
\affiliation{School of Physics, University of Western Australia, WA 6009, Australia}

\author{Andre N. Luiten}
\affiliation{Institute for Photonics and Advanced Sensing and School of Chemistry and Physics, University of Adelaide, SA 5005, Australia}
\affiliation{School of Physics, University of Western Australia, WA 6009, Australia}

\date{\today}

\begin{abstract}
We demonstrate thermometry with a resolution of 80 $\mathrm{nK} / \sqrt{\mathrm{Hz}}$ using  an isotropic crystalline whispering-gallery mode resonator based on a dichroic dual-mode technique. We simultaneously excite two modes that have a mode frequency ratio very close to two ($\pm0.3$ppm). The wavelength- and temperature-dependence of the refractive index means that the frequency difference between these modes is an ultra-sensitive proxy of the resonator temperature. This approach to temperature sensing automatically suppresses sensitivity to thermal expansion and vibrationally induced changes of the resonator.  We also demonstrate active suppression of temperature  fluctuations in the resonator by controlling  the intensity of the driving laser. The residual temperature fluctuations are shown to be below the limits set by fundamental thermodynamic fluctuations of the resonator material.
\begin{description}
\item[PACS numbers]
07.20.Dt, 42.60.Da, 42.62.Fi
\end{description}
\end{abstract}

\pacs{Valid PACS appear here}
\maketitle 


\noindent The high-resolution measurement of energy has long fascinated humans with its culmination seen in  ultra-high sensitivity calorimeters \cite{hansen2004art,wadso2002isothermal} and bolometers~\cite{richards1994bolometers}. These, and related ideas have found a broad range of applications including  bolometric superconducting photon-counters for quantum communication \cite{takesue2007quantum},  ultra-sensitive radio astronomy \cite{piat2003planck,woods2010demonstration}.  The  record for absolute thermometric sensitivity has been realised at cryogenic temperatures, achieving better than $100 \mathrm{pK} / \sqrt{\mathrm{Hz}}$ \cite{sergatskov2003new}.



In this letter we develop a new method to measure temperature based on excitation of a millimetre-scale Whispering-Gallery (WG) optical resonator with two widely frequency spaced modes.  These compact resonators  have  exceptionally high Q-factors and offer the potential to provide high-stability microwave and optical signals~\cite{mann2001cryogenic,tobar2006long,sprenger2010caf,liang2010whispering,alnis2011thermal}. Recently they been applied to high-sensitivity label-free sensors for molecules and viruses~\cite{armani2007label,Vollmer:2008} and for optical comb generation \cite{kippenberg2011microresonator}. Nonetheless, an issue that afflicts all these applications is the high temperature sensitivity of WG resonators~\cite{matsko2007whispering,alnis2011thermal}, particularly when compared to conventional vacuum-spaced Fabry-Perot resonators~\cite{kessler2012sub,amairi2012reducing,numata2004thermal,webster2008thermal,millo2009ultrastable}.  In this letter we turn this problem to our advantage by using the WG resonator  as an ultra-sensitive thermometer. 

To suppress unwanted temperature fluctuations in WG resonators several groups have demonstrated   \textit{in situ} thermometry  by measuring  the frequency difference between two orthogonally polarised modes. The best of these techniques have demonstrated a resolution of $\sim 100 \mathrm{nK} / \sqrt{\mathrm{Hz}}$ \cite{Strekalov:11}, and subsequent temperature stabilisation based on this sensing has resulted in improvement to the long term frequency stability \cite{fescenko2012dual,Baumgartel:12}. In  contrast, we present a two-colour approach to measure the resonator temperature with high resolution. In comparison to the birefringent dual-mode technique, our approach can be used in both anisotropic and isotropic resonators, which expands the range of material candidates.  Isotropic materials have shown the highest Q-factors to date \cite{savchenkov2007optical}, which offers potentially higher temperature resolution.  On the other hand, a combination of the dual-colour    and dual-polarisation approaches in an anisotropic material (e.g. MgF$_2$) can further enhance  the temperature sensitivity.  Furthermore, our dual-colour technique strongly rejects noise  from   thermal expansion fluctuations and vibrations, giving us the ability to   measure the mode-averaged temperature with a resolution below that of the fundamental thermal temperature fluctuations \cite{braginsky1999thermodynamical, gorodetsky2004fundamental, matsko2007whispering}.  

The frequency of a WG mode  depends on temperature through: (a) the temperature dependence of the refractive index (thermo-optic effect) as well as (b) the thermal expansion of the resonator. The first dependence leads to sensitivity to  the temperature solely within the  the optical mode, while in the latter the  the mode frequency depends on the temperature distribution throughout the entire resonator volume. For simplicity, we assume  a steady-state temperature distribution, $T_{P}(r)$, when the resonator is excited by some  input optical power, $P$, that is solely dependent on the radial co-ordinate, $r$. This approximation reflects  the typical  triple cylindrical symmetry exhibited by the (i) resonator geometry,  (ii) optical power distribution and (iii) thermal coupling to the external environment.  When the power-induced temperature changes from ambient are small (i.e. $\Delta T_{P}(r) = T_{P}(r)-T_{0}(r) \ll T_{0}(r)$), we can express the frequency, $f_m$, of the $m^\textrm{th}$ mode as:

\begin{equation}
\frac{ f_{m}}{f_{m,0}}= 1-\frac{2  \alpha \int_{0}^{R} \Delta T_{P}(r) \, r dr}{R^2}-\frac{\beta(f_m)}{n(f_m)}\Delta T_{P}(R)-\gamma_m
 \label{modefreq}
\end{equation}
where $f_{m,0}$ is the frequency of the $m^\textrm{th}$ mode in the absence of excitation power, $\alpha$ is the linear thermal expansion coefficient,  and $\beta(f_m)$ and $n(f_m)$ are the  thermo-optic coefficient and refractive index of the resonator material, respectively,  $R$ is the radius at the mode intensity maximum, 
and  $\gamma_m=\frac{2 Q_0}{Q_0+Q_c} \frac{n_{Kerr}}{n(f_m)} \frac{\cal F}{\pi} \frac{P}{A_m}$ characterises the refractive index dependence upon the optical intensity, which depends on the Kerr coefficient $n_{Kerr}$, and finesse $\cal F$ \cite{schliesser}. $Q_0$ and $Q_c$ are the intrinsic Q-factor and coupling Q-factor respectively. We define an effective mode area $A_m=\frac{\int |E_m|^2 dA \int |E_{tot}|^2 dA}{\int |E_m|^2 |E_{tot}|^2 dA}$, where $E_{m}$ is the  amplitude of the $m^\textrm{th}$ mode in the transverse plane and $E_{tot}$ is the  field amplitude of the resonant energy (which allows for more than 1 mode to be excited simultaneously). 

The basis  of our thermometer is to simultaneously lock two optical signals to two WG modes that have frequencies  $f_{1}$ and $f_{2}$ with $ f_{2} \approx 2 f_{1}$. These modes are chosen to be within the same transverse mode family (i.e identical polar and radial field maxima numbers \cite{little1999analytic}) to maximise their spatial overlap.
For simplicity  the two optical signals are derived from a single laser  with frequency $f_{L}$.  

The direct output of the laser is locked using the Pound-Drever-Hall (PDH) technique \cite{drever1983laser} to the lower frequency WG mode, i.e.\ $f_L=f_1$. We   frequency double this laser signal and shift it into resonance with the second WG mode using an Acousto-Optic Modulator (AOM), so that $2 f_{L}+f_{AOM} = f_{2}$.
From Eq.~\ref{modefreq}, it follows that 
\begin{equation}
f_{AOM}
\approx 2 f_{L} \left(\frac{\beta(f_1)}{n(f_1)}-\frac{\beta(f_2)}{n(f_2)}\right) \Delta T_{P}(R)+C+\Gamma+N
\label{aomfreq}
\end{equation}
where $C=f_{2,0}-2f_{1,0}$,  $\Gamma =  2 f_{L} (\gamma_1-\gamma_2)$ is the relative non-linear Kerr shift, and $N=\delta f_{2}-2 \delta f_{1}$ which accounts for residual errors in the laser locking systems, i.e $\delta f_1=f_L-f_1$ and $\delta f_2=2 f_{L}+ f_{AOM}-f_2$. Importantly,  the high degree of spatial overlap between the two modes gives rise to nearly identical frequency dependence on thermal expansion, so that the distributed  thermal expansion term appearing in Eq.~\ref{modefreq} is strongly suppressed in Eq.~\ref{aomfreq}; we estimate its fractional contribution to be less than 1 part in  $10^{6}$ and so we ignore it in what follows.  

The first term on right hand side in Eq.~\ref{aomfreq} indicates that the AOM frequency  provides a high-quality read-out of the resonator temperature  if the thermo-optic coefficient at $f_{1}$ and $f_{2}$ is sufficiently different to dominate over the noise terms, $\Gamma + N$. We show that this is the case for this resonator with sensing in the nK regime \footnote{
We note that suppressing mode-averaged temperature fluctuations using this technique does not significantly improve the absolute frequency stability of the WGM resonator.   Eq.~\ref{modefreq} shows that the mode frequency depends on the temperature distribution throughout the resonator, which is not fixed by mode-averaged temperature control \cite{fescenko2012dual,Baumgartel:12}. }.

\begin{figure}[htb]
\centerline{\includegraphics[width=8.5cm]{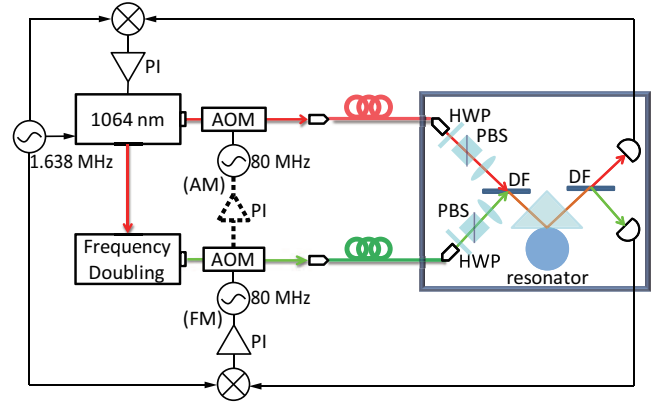}}
\caption{Experimental setup. HWP: half-wave plate; PBS: polarising beam splitter; DF: dichroic filter; AM: amplitude modulation; FM: frequency modulation. The dashed PI control part was only implemented when stabilising the temperature of the resonator, i.e. it was not included in the temperature sensing experiments.}
\label{setup}
\end{figure}

The experimental setup is shown in Fig.~\ref{setup}. A 5\,mm radius CaF$_2$ WGM resonator is mounted on a piezo-actuated translation stage within  an acoustic and thermal shield.  The laser light was coupled into the resonator using a high index prism. A conventional thermometer/heater pair was used to pre-stabilise the temperature of the system at $\pm$0.1\,K level.  Light from a Nd:YAG laser at 1064\,nm, together with its second harmonic at 532\,nm (generated in a  single-pass nonlinear crystal),  were transferred into the shielded volume using single-mode optical fibres. The 532\,nm light was double-passed through an AOM to enable independent frequency tuning of this beam. The 1064\,nm light was also double-passed through a second AOM for reasons explained below. The two AOMs were driven with independent oscillators with nominal frequency around 80\,MHz and were set to shift the frequency upwards. Both beams were recombined inside the shielded volume using a dichroic filter before being coupled into the resonator. The transmitted beams were separated using a second dichroic filter and then registered by two photodetectors. The laser was frequency modulated at 1.638 MHz and the two detected signals were synchronously demodulated using the traditional PDH technique to generate independent error signals appropriate to lock the laser signals onto their respective modes.  The error signal generated from the 1064\,nm mode was integrated and sent directly back to the laser controller to maintain the frequency lock. The second harmonic light  was frequency  locked by controlling the frequency of the synthesiser that drove the AOM. 

 As the evanescent field has different scale lengths for the two modes \cite{little1999analytic},  it was necessary to over-couple the 1064\,nm mode in order to achieve adequate coupling for the 532\,nm mode. Thus, the loaded Q for the modes was  $2.6\times 10^8$ and  $3.6\times 10^8$ for the 1064\,nm and 532\,nm modes respectively. This situation may be overcome by designing an appropriate coupling scheme \cite{Ghulinyan2013oscillatory}.
 
The fast fluctuations of the WG-resonator thermometer were monitored by observing the frequency fluctuations of $f_{AOM}$ with a spectrum analyser, while slower fluctuations were monitored by a frequency counter. In addition, we  measured the frequency of the lower frequency mode ($f_1$) with a stabilised frequency comb ($\Delta f/f < 10^{-13}$ for time scales  $>$ 1 s). For these experiments, the excitation power was deliberately kept at low levels (50 $\mu$W for 532\,nm and 70\,$\mu$W for 1064\,nm) so that photo-thermal  \cite{matsko2007whispering,goda2005photothermal} and Kerr noise were both at least 10\,dB below the measured $f_{AOM}$ spectrum at all frequencies. The low levels of these noise sources during  operation  was verified by increasing the  input power until both effects were observed and then  turning the power back down by a factor of ten when in operation. 

\begin{figure}
\centerline{\includegraphics[width=7.5cm]{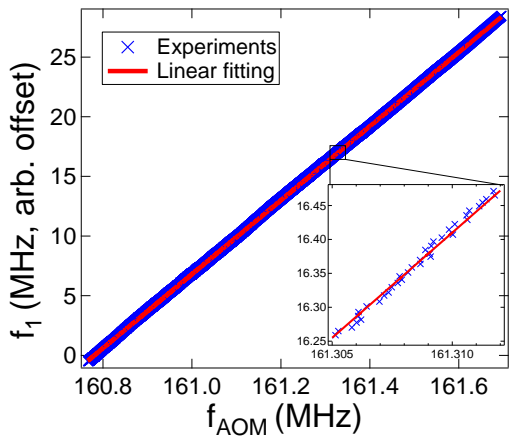}}
\caption{Relation between $f_1$ and $f_{AOM}$ while the resonator temperature drifts over 10 mK.   Inset shows an enlarged fraction of the data. De-correlation  arises because  the mode frequency depends on both the mode-averaged temperature as well as the entire temperature distribution, whereas the thermometer is only sensitive to the mode-averaged temperature.}
\label{freerun}
\end{figure}

Fig.~\ref{freerun} displays the variation of the lower mode frequency ($f_1$) as a function of the frequency of the AOM (a proxy for the mode temperature) where the resonator temperature drifts by 10\,mK. As expected, there is a very high degree of correlation between $f_1$ and $f_{AOM}$ (see inset), with  slope  $\Delta f_1/\Delta f_{AOM}=31.0$. We  calculate the expected mode-frequency sensitivity from the known parameter values for CaF$_{2}$ at 1064\,nm ($n=1.43$, $\beta=-11.4\times 10^{-6}$/K and $\alpha=18.7\times 10^{-6}$/K at 1064 nm \cite{savchenkov2007whispering,feldman1979optical})  as $d f_1/d T=-3.02$ GHz/K. We were able to verify this calculated value at the level of 10\% by changing the temperature of the resonator mount intentionally.  The   mode-frequency to temperature relation is  combined with the observed relation to the AOM frequency to give a thermometer calibration of  $d f_{AOM}/d T=-97.42$ MHz/K.   In what follows, we examine the  fluctuations of $f_{AOM}$ in more detail, but first we explain the means for temperature control of the resonator.

\begin{figure}
\centerline{\includegraphics[width=8.5cm]{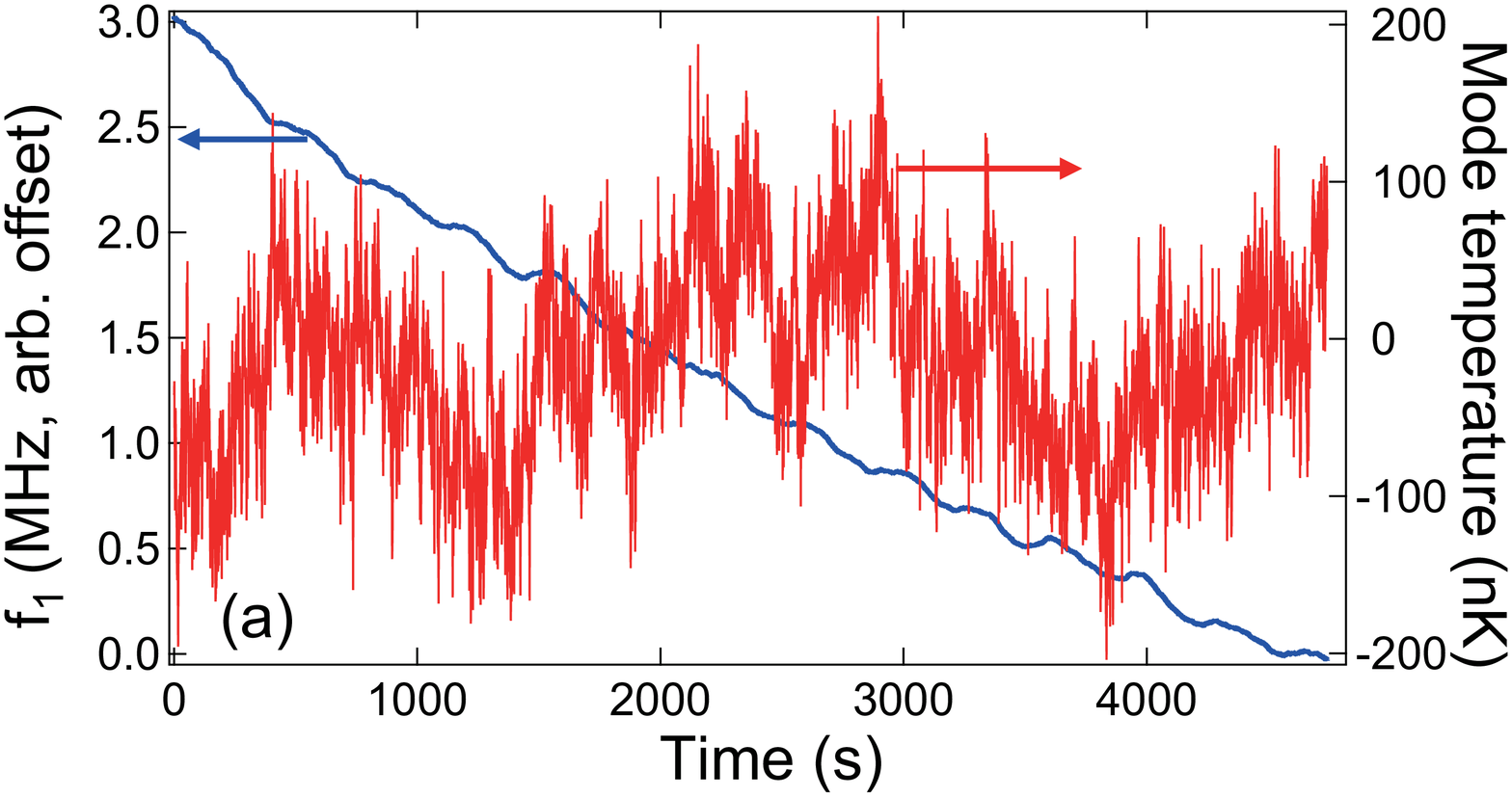}}
\centerline{\includegraphics[width=9cm]{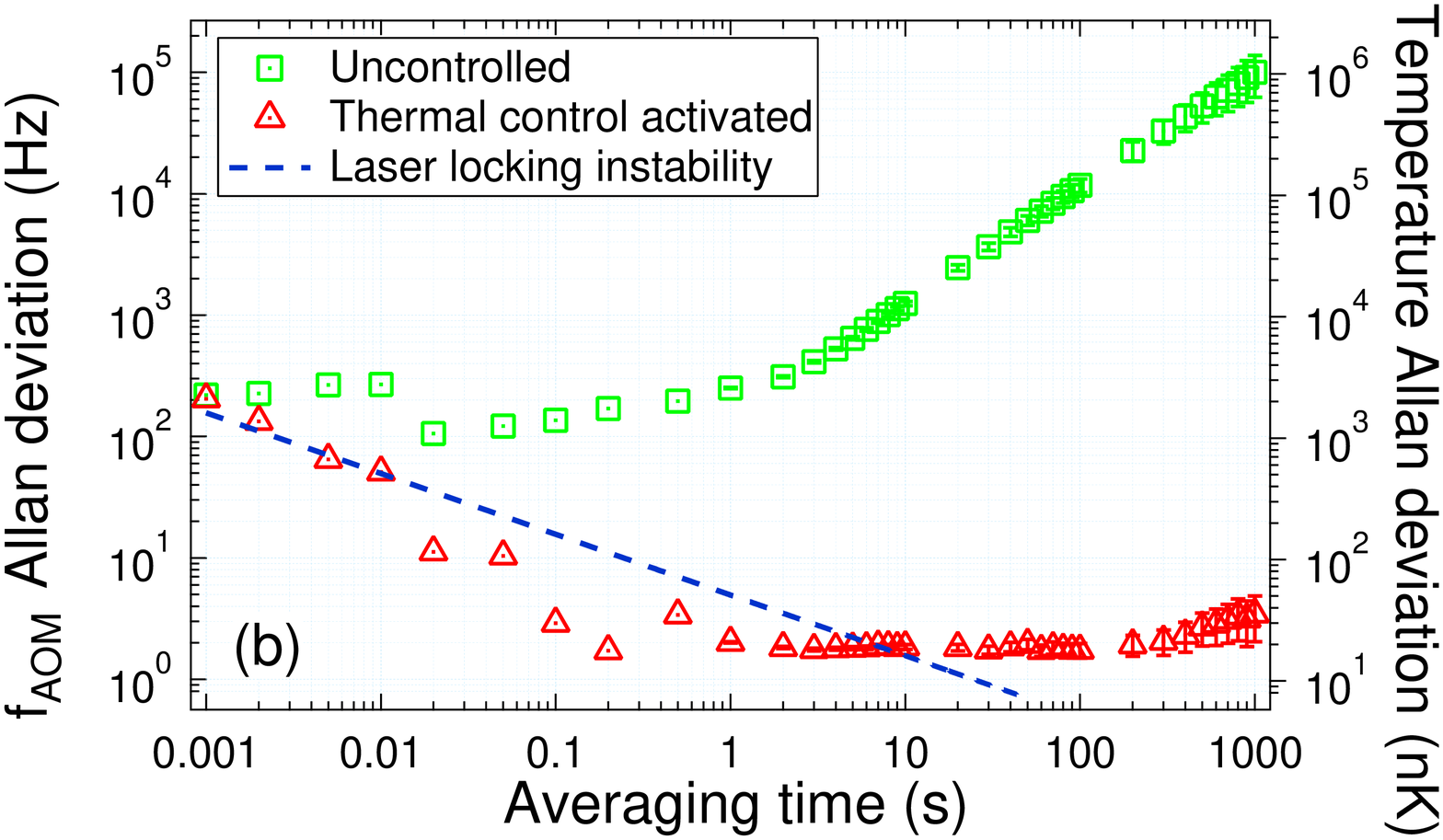}}
\centerline{\includegraphics[width=8.8cm]{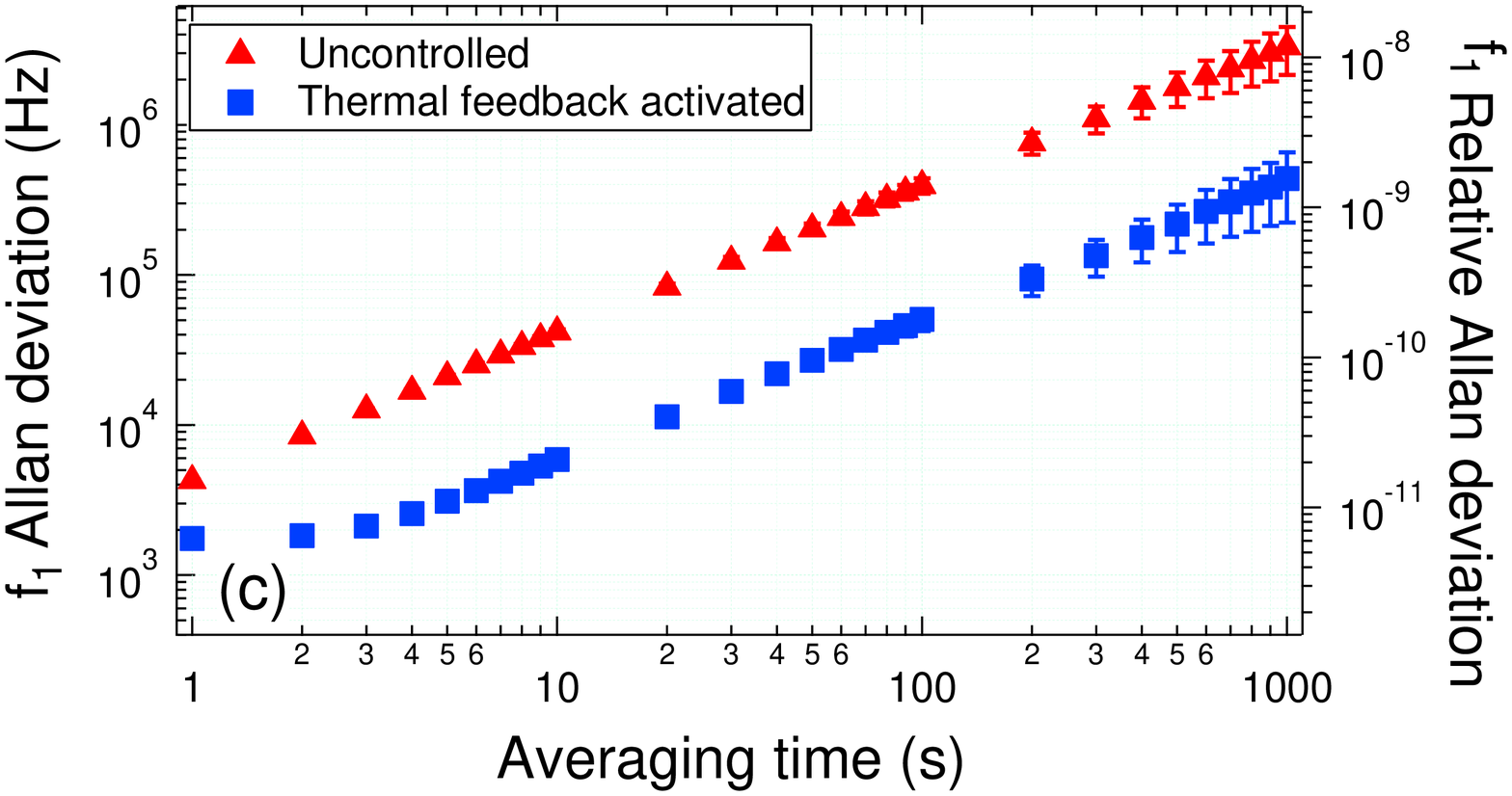}}
\caption{(a) Mode frequency ($f_1$) and the mode temperature derived from $f_{AOM}$ with thermal feedback control activated. (b) Allan deviations of $f_{AOM}$ and the imputed mode temperature when temperature is uncontrolled (green squares) and actively controlled (red triangles). Dashed blue line is the estimated laser locking instability which limits the achievable mode volume temperature stability. (c) Allan deviation of the 1064 nm signal when the mode volume temperature is uncontrolled (red triangles) and actively controlled (blue squares).}
\label{controlled}
\end{figure}

The resonator temperature can be controlled with  a high bandwidth by actively controlling the input optical power.  For these experiments we increased the 1064\,nm power to $\sim 2$\,mW to increase the range of the temperature control system. By directly measuring the transfer function between $ f_{AOM}$  and the 1064\,nm power, we find the control bandwidth is $>1$\,kHz.  The control actuator is indicated on Fig.~\ref{setup} as the dotted line in which  the drive power of the 1064\,nm AOM is actively controlled to maintain the frequency of the AOM at a fixed value, which locks the resonator temperature. With this thermal control, the average mode temperature can be stabilised at the 100 nK level for more than an hour as seen on Fig.~\ref{controlled}(a).  A residual drift of $\sim 2.5$\,MHz/hr in the mode frequency arises because it depends upon the entire temperature distribution, which is uncontrolled. The ripples with a $\sim 300$\,s period are associated with room temperature modulation from the air-conditioning.  Fig.\ref{controlled}(b) shows a time-domain representation of the temperature fluctuations of the controlled and uncontrolled resonator using the Allan deviation \cite{allan1966statistics}. We see that the control system suppresses the long-term temperature fluctuations by more than 4 orders of magnitude to the 30\,nK level. The long term temperature stability appears flat as a result of the interaction of the free running fluctuations and the transfer function of our control loop: a more sophisticated control system could result in further suppression. In Fig.~\ref{controlled}(c), we show the Allan deviation of the locked 1064 nm mode frequency when the temperature is stabilised. The performance is substantially worse than one would expect if the mode frequency only depended upon the temperature in the mode ($\sim5\times10^{-13}$ at 1 second averaging time). Nonetheless, the temperature control technique suppressed the mode frequency fluctuations by nearly 1 order of magnitude.

\begin{figure}
\centerline{\includegraphics[width=8.8cm]{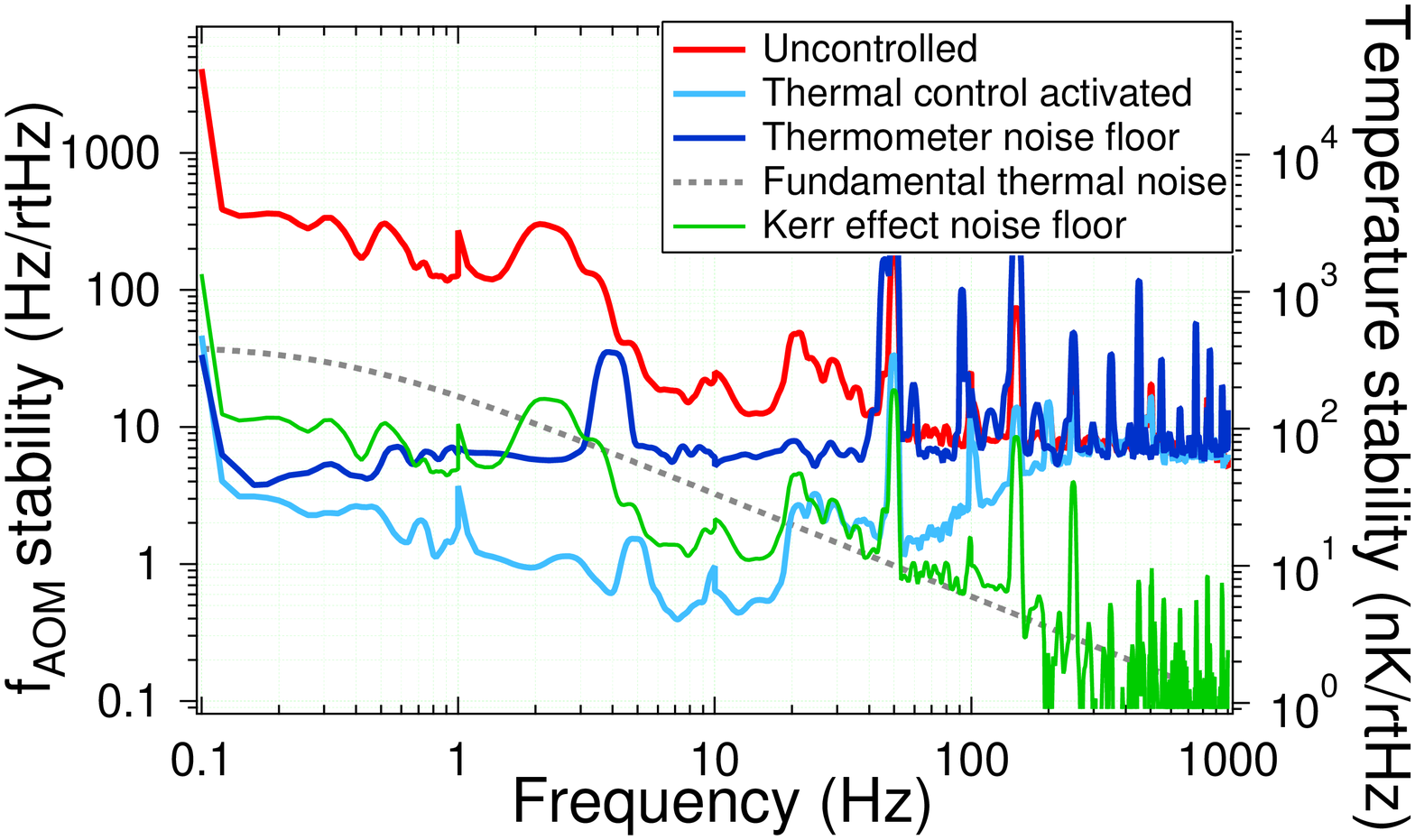}}
\caption{PSD of $f_{AOM}$ and its projected mode temperature. $f_{AOM}$ is an in-loop signal that contains the lock instability noise, so below the lock instability noise floor (dark blue trace) the stabilised $f_{AOM}$ PSD (light blue trace) does not project the real temperature PSD.}
\label{fancy}
\end{figure}

Fig.~\ref{fancy} shows the power spectral density (PSD) of $f_{AOM}$ when it is  free-running, and when it is actively controlled.  We also display the fundamental temperature fluctuations calculated using the method in \cite{matsko2007whispering}.  We   assume that the two optical modes have large spatial overlap, which is reasonable given that the thermal wavelength of even the highest frequency thermal noise ($\sim 1$\,kHz) considered here is much larger than the transverse extent of the optical modes, or their separation \cite{matsko2007whispering}. The unstabilised mode temperature fluctuation exceeds the calculated fundamental thermal fluctuations by a factor of 10 at low frequencies because the resonator is also subject to fluctuations in ambient temperature and input-power. 

Fig.~\ref{fancy} also shows the noise floor of the temperature sensor (i.e.\ $N+\Gamma$ in Eq.~\ref{aomfreq}), which arises from the residual frequency noise  in the two optical frequency stabilisation loops as well as the effect of Kerr fluctuations induced by the input intensity control. Residual frequency noise in the stabilisation loops was independently estimated by measuring the noise in the frequency locking systems  when the lasers were detuned from resonance: in these circumstances, we measure the sum of any electronic, shot-noise, residual amplitude modulation (RAM) and residual intensity noise (RIN) that limit the frequency stabilization loops. The RIN in the environs of the PDH modulation frequency was seen to be the dominant contributor to this noise limit and sets a resulting 80 $\mathrm{nK} / \sqrt{\mathrm{Hz}}$ temperature sensitivity that is reasonably frequency-independent.   It can be seen that this sensitivity is below the fundamental thermal noise of the resonator for thermal frequencies below 3\,Hz.  

Finally,  Fig.~\ref{fancy} shows the  residual temperature fluctuations once the mode temperature is stabilised: the fluctuations are strongly suppressed by the control system within its 200\,Hz thermal control bandwidth. 

We note that an additional noise floor arises when the resonator temperature is actively controlled: the required intensity modulation causes frequency shifts both through the desired temperature changes but also through an unwanted Kerr effect. This arises because  the self-mode Kerr shift and cross-Kerr mode shift differ because of the differing   cross-sectional areas for the two modes. This type of noise floor could be minimised by modulating the power of both excited modes together in a judiciously chosen ratio that results in the same effective Kerr shift in the two modes.  We have not undertaken this procedure here since it was not the limiting factor in the performance of the temperature sensor.  This induced noise floor is also shown on  Fig.~\ref{fancy} and was determined by measuring the  spectrum of the  intensity modulation required to keep the temperature stabilised. The transfer coefficient between intensity and the resulting Kerr frequency shift was measured by applying an intentionally large intensity modulation. The resulting Kerr sensitivity was in broad agreement with the theoretical shifts expected from the relatively large mode volumes ($\sim 100 \mu \mathrm{m} \times   2 \mu \mathrm{m} \times 2$\,cm).

This experiment demonstrates the potential of the technique.  There are several routes to further improve the  sensitivity and resolution: (a) Reduce the residual noise in the frequency locking system by increasing the Q-factor of the modes. To this end, a more sophisticated coupling would reduce the current difference in evanescent coupling strengths, thereby improving the Q-factor \cite{Ghulinyan2013oscillatory}. (b) Increase the difference between the thermo-optic coefficients for the fundamental and second harmonic by moving to fundamental wavelength of 800-900\,nm. This difference can be $\sim4$ times larger than that of the two wavelengths used here \cite{ghosh1998handbook}. (c) Combine the wavelength- and polarisation-dependent thermal-sensing techniques  in a material such as MgF$_2$. Combining these approaches, we estimate that it is feasible to obtain a sensitivity below 10 $\mathrm{nK} / \sqrt{\mathrm{Hz}}$  with the same detection noise.  

To conclude, we report a dichroic mode temperature sensing and thermal stabilisation scheme in a WGM resonator.  The experiment  achieves tens of nano-Kelvin temperature stability by suppressing thermal fluctuations below the fundamental thermal noise level. This technique opens the possibility for  the WG mode resonator as an ultra-sensitive thermometer.

\textbf{Acknowledgements}
The authors gratefully acknowledge financial support from the Australian Research Council under grants DP0877938, FT0991631, LE110100054 and LE100100009 that enabled this work. The authors also wish to acknowledge the South Australian Government who have provided generous financial support through the Premier's Science and Research Fund. The authors thank Ping Koy Lam for developing the facility for shaping the resonator and Frank Van Kann for useful discussions. W. Weng gratefully acknowledges financial support from China Scholarship Council and University of Western Australia.

\bibliography{Ref}

\begin{thebibliography}{37}%
\makeatletter
\providecommand \@ifxundefined [1]{%
 \@ifx{#1\undefined}
}%
\providecommand \@ifnum [1]{%
 \ifnum #1\expandafter \@firstoftwo
 \else \expandafter \@secondoftwo
 \fi
}%
\providecommand \@ifx [1]{%
 \ifx #1\expandafter \@firstoftwo
 \else \expandafter \@secondoftwo
 \fi
}%
\providecommand \natexlab [1]{#1}%
\providecommand \enquote  [1]{``#1''}%
\providecommand \bibnamefont  [1]{#1}%
\providecommand \bibfnamefont [1]{#1}%
\providecommand \citenamefont [1]{#1}%
\providecommand \href@noop [0]{\@secondoftwo}%
\providecommand \href [0]{\begingroup \@sanitize@url \@href}%
\providecommand \@href[1]{\@@startlink{#1}\@@href}%
\providecommand \@@href[1]{\endgroup#1\@@endlink}%
\providecommand \@sanitize@url [0]{\catcode `\\12\catcode `\$12\catcode
  `\&12\catcode `\#12\catcode `\^12\catcode `\_12\catcode `\%12\relax}%
\providecommand \@@startlink[1]{}%
\providecommand \@@endlink[0]{}%
\providecommand \url  [0]{\begingroup\@sanitize@url \@url }%
\providecommand \@url [1]{\endgroup\@href {#1}{\urlprefix }}%
\providecommand \urlprefix  [0]{URL }%
\providecommand \Eprint [0]{\href }%
\providecommand \doibase [0]{http://dx.doi.org/}%
\providecommand \selectlanguage [0]{\@gobble}%
\providecommand \bibinfo  [0]{\@secondoftwo}%
\providecommand \bibfield  [0]{\@secondoftwo}%
\providecommand \translation [1]{[#1]}%
\providecommand \BibitemOpen [0]{}%
\providecommand \bibitemStop [0]{}%
\providecommand \bibitemNoStop [0]{.\EOS\space}%
\providecommand \EOS [0]{\spacefactor3000\relax}%
\providecommand \BibitemShut  [1]{\csname bibitem#1\endcsname}%
\let\auto@bib@innerbib\@empty
\bibitem [{\citenamefont {Hansen}\ and\ \citenamefont
  {Hart}(2004)}]{hansen2004art}%
  \BibitemOpen
  \bibfield  {author} {\bibinfo {author} {\bibfnamefont {L.~D.}\ \bibnamefont
  {Hansen}}\ and\ \bibinfo {author} {\bibfnamefont {R.~M.}\ \bibnamefont
  {Hart}},\ }\href@noop {} {\bibfield  {journal} {\bibinfo  {journal}
  {Thermochim. Acta}\ }\textbf {\bibinfo {volume} {417}},\ \bibinfo {pages}
  {257} (\bibinfo {year} {2004})}\BibitemShut {NoStop}%
\bibitem [{\citenamefont {Wads{\"o}}(2002)}]{wadso2002isothermal}%
  \BibitemOpen
  \bibfield  {author} {\bibinfo {author} {\bibfnamefont {I.}~\bibnamefont
  {Wads{\"o}}},\ }\href@noop {} {\bibfield  {journal} {\bibinfo  {journal}
  {Thermochim. Acta}\ }\textbf {\bibinfo {volume} {394}},\ \bibinfo {pages}
  {305} (\bibinfo {year} {2002})}\BibitemShut {NoStop}%
\bibitem [{\citenamefont {Richards}(1994)}]{richards1994bolometers}%
  \BibitemOpen
  \bibfield  {author} {\bibinfo {author} {\bibfnamefont {P.}~\bibnamefont
  {Richards}},\ }\href@noop {} {\bibfield  {journal} {\bibinfo  {journal} {J.
  Appl. Phys.}\ }\textbf {\bibinfo {volume} {76}},\ \bibinfo {pages} {1}
  (\bibinfo {year} {1994})}\BibitemShut {NoStop}%
\bibitem [{\citenamefont {Takesue}\ \emph {et~al.}(2007)\citenamefont
  {Takesue}, \citenamefont {Nam}, \citenamefont {Zhang}, \citenamefont
  {Hadfield}, \citenamefont {Honjo}, \citenamefont {Tamaki},\ and\
  \citenamefont {Yamamoto}}]{takesue2007quantum}%
  \BibitemOpen
  \bibfield  {author} {\bibinfo {author} {\bibfnamefont {H.}~\bibnamefont
  {Takesue}}, \bibinfo {author} {\bibfnamefont {S.~W.}\ \bibnamefont {Nam}},
  \bibinfo {author} {\bibfnamefont {Q.}~\bibnamefont {Zhang}}, \bibinfo
  {author} {\bibfnamefont {R.~H.}\ \bibnamefont {Hadfield}}, \bibinfo {author}
  {\bibfnamefont {T.}~\bibnamefont {Honjo}}, \bibinfo {author} {\bibfnamefont
  {K.}~\bibnamefont {Tamaki}}, \ and\ \bibinfo {author} {\bibfnamefont
  {Y.}~\bibnamefont {Yamamoto}},\ }\href@noop {} {\bibfield  {journal}
  {\bibinfo  {journal} {Nature photonics}\ }\textbf {\bibinfo {volume} {1}},\
  \bibinfo {pages} {343} (\bibinfo {year} {2007})}\BibitemShut {NoStop}%
\bibitem [{\citenamefont {Piat}\ \emph {et~al.}(2003)\citenamefont {Piat},
  \citenamefont {Lamarre}, \citenamefont {Meissonnier}, \citenamefont {Torre},
  \citenamefont {Camus}, \citenamefont {Benoit}, \citenamefont {Crussaire},
  \citenamefont {Ade}, \citenamefont {Bock}, \citenamefont {Lange} \emph
  {et~al.}}]{piat2003planck}%
  \BibitemOpen
  \bibfield  {author} {\bibinfo {author} {\bibfnamefont {M.}~\bibnamefont
  {Piat}}, \bibinfo {author} {\bibfnamefont {J.-M.}\ \bibnamefont {Lamarre}},
  \bibinfo {author} {\bibfnamefont {J.}~\bibnamefont {Meissonnier}}, \bibinfo
  {author} {\bibfnamefont {J.-P.}\ \bibnamefont {Torre}}, \bibinfo {author}
  {\bibfnamefont {P.}~\bibnamefont {Camus}}, \bibinfo {author} {\bibfnamefont
  {A.}~\bibnamefont {Benoit}}, \bibinfo {author} {\bibfnamefont {J.-P.}\
  \bibnamefont {Crussaire}}, \bibinfo {author} {\bibfnamefont {P.~A.}\
  \bibnamefont {Ade}}, \bibinfo {author} {\bibfnamefont {J.~J.}\ \bibnamefont
  {Bock}}, \bibinfo {author} {\bibfnamefont {A.~E.}\ \bibnamefont {Lange}},
  \emph {et~al.},\ }in\ \href@noop {} {\emph {\bibinfo {booktitle}
  {Astronomical Telescopes and Instrumentation}}}\ (\bibinfo {organization}
  {International Society for Optics and Photonics},\ \bibinfo {year} {2003})\
  pp.\ \bibinfo {pages} {740--748}\BibitemShut {NoStop}%
\bibitem [{\citenamefont {Woods}\ \emph {et~al.}(2010)\citenamefont {Woods},
  \citenamefont {Carr}, \citenamefont {Jung}, \citenamefont {Carter},\ and\
  \citenamefont {Datla}}]{woods2010demonstration}%
  \BibitemOpen
  \bibfield  {author} {\bibinfo {author} {\bibfnamefont {S.~I.}\ \bibnamefont
  {Woods}}, \bibinfo {author} {\bibfnamefont {S.~M.}\ \bibnamefont {Carr}},
  \bibinfo {author} {\bibfnamefont {T.~M.}\ \bibnamefont {Jung}}, \bibinfo
  {author} {\bibfnamefont {A.~C.}\ \bibnamefont {Carter}}, \ and\ \bibinfo
  {author} {\bibfnamefont {R.~U.}\ \bibnamefont {Datla}},\ }\href {\doibase
  http://dx.doi.org/10.1063/1.3456539} {\bibfield  {journal} {\bibinfo
  {journal} {Journal of Applied Physics}\ }\textbf {\bibinfo {volume} {108}},\
  \bibinfo {eid} {024505} (\bibinfo {year} {2010})}\BibitemShut {NoStop}%
\bibitem [{\citenamefont {Sergatskov}\ \emph {et~al.}(2003)\citenamefont
  {Sergatskov}, \citenamefont {Day}, \citenamefont {Babkin}, \citenamefont
  {Nelson}, \citenamefont {McCarson}, \citenamefont {Boyd},\ and\ \citenamefont
  {Duncan}}]{sergatskov2003new}%
  \BibitemOpen
  \bibfield  {author} {\bibinfo {author} {\bibfnamefont {D.~A.}\ \bibnamefont
  {Sergatskov}}, \bibinfo {author} {\bibfnamefont {P.~K.}\ \bibnamefont {Day}},
  \bibinfo {author} {\bibfnamefont {A.~V.}\ \bibnamefont {Babkin}}, \bibinfo
  {author} {\bibfnamefont {R.~C.}\ \bibnamefont {Nelson}}, \bibinfo {author}
  {\bibfnamefont {T.~D.}\ \bibnamefont {McCarson}}, \bibinfo {author}
  {\bibfnamefont {S.~T.~P.}\ \bibnamefont {Boyd}}, \ and\ \bibinfo {author}
  {\bibfnamefont {R.~V.}\ \bibnamefont {Duncan}},\ }\href@noop {} {\bibfield
  {journal} {\bibinfo  {journal} {AIP Conference Proceedings}\ }\textbf
  {\bibinfo {volume} {684}},\ \bibinfo {pages} {1009} (\bibinfo {year}
  {2003})}\BibitemShut {NoStop}%
\bibitem [{\citenamefont {Mann}\ \emph {et~al.}(2001)\citenamefont {Mann},
  \citenamefont {Sheng},\ and\ \citenamefont {Luiten}}]{mann2001cryogenic}%
  \BibitemOpen
  \bibfield  {author} {\bibinfo {author} {\bibfnamefont {A.~G.}\ \bibnamefont
  {Mann}}, \bibinfo {author} {\bibfnamefont {C.}~\bibnamefont {Sheng}}, \ and\
  \bibinfo {author} {\bibfnamefont {A.~N.}\ \bibnamefont {Luiten}},\
  }\href@noop {} {\bibfield  {journal} {\bibinfo  {journal} {IEEE Trans.
  Instrum. Meas.}\ }\textbf {\bibinfo {volume} {50}},\ \bibinfo {pages} {519}
  (\bibinfo {year} {2001})}\BibitemShut {NoStop}%
\bibitem [{\citenamefont {Tobar}\ \emph {et~al.}(2006)\citenamefont {Tobar},
  \citenamefont {Ivanov}, \citenamefont {Locke}, \citenamefont {Stanwix},
  \citenamefont {Hartnett}, \citenamefont {Luiten}, \citenamefont {Warrington},
  \citenamefont {Fisk}, \citenamefont {Lawn}, \citenamefont {Wouters} \emph
  {et~al.}}]{tobar2006long}%
  \BibitemOpen
  \bibfield  {author} {\bibinfo {author} {\bibfnamefont {M.~E.}\ \bibnamefont
  {Tobar}}, \bibinfo {author} {\bibfnamefont {E.~N.}\ \bibnamefont {Ivanov}},
  \bibinfo {author} {\bibfnamefont {C.~R.}\ \bibnamefont {Locke}}, \bibinfo
  {author} {\bibfnamefont {P.~L.}\ \bibnamefont {Stanwix}}, \bibinfo {author}
  {\bibfnamefont {J.~G.}\ \bibnamefont {Hartnett}}, \bibinfo {author}
  {\bibfnamefont {A.~N.}\ \bibnamefont {Luiten}}, \bibinfo {author}
  {\bibfnamefont {R.~B.}\ \bibnamefont {Warrington}}, \bibinfo {author}
  {\bibfnamefont {P.~T.}\ \bibnamefont {Fisk}}, \bibinfo {author}
  {\bibfnamefont {M.~A.}\ \bibnamefont {Lawn}}, \bibinfo {author}
  {\bibfnamefont {M.~J.}\ \bibnamefont {Wouters}},  \emph {et~al.},\
  }\href@noop {} {\bibfield  {journal} {\bibinfo  {journal} {IEEE Trans.
  Ultrason., Ferroelectr. Freq. Control}\ }\textbf {\bibinfo {volume} {53}},\
  \bibinfo {pages} {2386} (\bibinfo {year} {2006})}\BibitemShut {NoStop}%
\bibitem [{\citenamefont {Sprenger}\ \emph {et~al.}(2010)\citenamefont
  {Sprenger}, \citenamefont {Schwefel}, \citenamefont {Lu}, \citenamefont
  {Svitlov},\ and\ \citenamefont {Wang}}]{sprenger2010caf}%
  \BibitemOpen
  \bibfield  {author} {\bibinfo {author} {\bibfnamefont {B.}~\bibnamefont
  {Sprenger}}, \bibinfo {author} {\bibfnamefont {H.}~\bibnamefont {Schwefel}},
  \bibinfo {author} {\bibfnamefont {Z.}~\bibnamefont {Lu}}, \bibinfo {author}
  {\bibfnamefont {S.}~\bibnamefont {Svitlov}}, \ and\ \bibinfo {author}
  {\bibfnamefont {L.}~\bibnamefont {Wang}},\ }\href@noop {} {\bibfield
  {journal} {\bibinfo  {journal} {Opt. Lett.}\ }\textbf {\bibinfo {volume}
  {35}},\ \bibinfo {pages} {2870} (\bibinfo {year} {2010})}\BibitemShut
  {NoStop}%
\bibitem [{\citenamefont {Liang}\ \emph {et~al.}(2010)\citenamefont {Liang},
  \citenamefont {Ilchenko}, \citenamefont {Savchenkov}, \citenamefont {Matsko},
  \citenamefont {Seidel},\ and\ \citenamefont {Maleki}}]{liang2010whispering}%
  \BibitemOpen
  \bibfield  {author} {\bibinfo {author} {\bibfnamefont {W.}~\bibnamefont
  {Liang}}, \bibinfo {author} {\bibfnamefont {V.}~\bibnamefont {Ilchenko}},
  \bibinfo {author} {\bibfnamefont {A.}~\bibnamefont {Savchenkov}}, \bibinfo
  {author} {\bibfnamefont {A.}~\bibnamefont {Matsko}}, \bibinfo {author}
  {\bibfnamefont {D.}~\bibnamefont {Seidel}}, \ and\ \bibinfo {author}
  {\bibfnamefont {L.}~\bibnamefont {Maleki}},\ }\href@noop {} {\bibfield
  {journal} {\bibinfo  {journal} {Opt. Lett.}\ }\textbf {\bibinfo {volume}
  {35}},\ \bibinfo {pages} {2822} (\bibinfo {year} {2010})}\BibitemShut
  {NoStop}%
\bibitem [{\citenamefont {Alnis}\ \emph {et~al.}(2011)\citenamefont {Alnis},
  \citenamefont {Schliesser}, \citenamefont {Wang}, \citenamefont {Hofer},
  \citenamefont {Kippenberg},\ and\ \citenamefont
  {H{\"a}nsch}}]{alnis2011thermal}%
  \BibitemOpen
  \bibfield  {author} {\bibinfo {author} {\bibfnamefont {J.}~\bibnamefont
  {Alnis}}, \bibinfo {author} {\bibfnamefont {A.}~\bibnamefont {Schliesser}},
  \bibinfo {author} {\bibfnamefont {C.~Y.}\ \bibnamefont {Wang}}, \bibinfo
  {author} {\bibfnamefont {J.}~\bibnamefont {Hofer}}, \bibinfo {author}
  {\bibfnamefont {T.~J.}\ \bibnamefont {Kippenberg}}, \ and\ \bibinfo {author}
  {\bibfnamefont {T.~W.}\ \bibnamefont {H{\"a}nsch}},\ }\href@noop {}
  {\bibfield  {journal} {\bibinfo  {journal} {Phys. Rev. A}\ }\textbf {\bibinfo
  {volume} {84}},\ \bibinfo {pages} {011804} (\bibinfo {year}
  {2011})}\BibitemShut {NoStop}%
\bibitem [{\citenamefont {Armani}\ \emph {et~al.}(2007)\citenamefont {Armani},
  \citenamefont {Kulkarni}, \citenamefont {Fraser}, \citenamefont {Flagan},\
  and\ \citenamefont {Vahala}}]{armani2007label}%
  \BibitemOpen
  \bibfield  {author} {\bibinfo {author} {\bibfnamefont {A.~M.}\ \bibnamefont
  {Armani}}, \bibinfo {author} {\bibfnamefont {R.~P.}\ \bibnamefont
  {Kulkarni}}, \bibinfo {author} {\bibfnamefont {S.~E.}\ \bibnamefont
  {Fraser}}, \bibinfo {author} {\bibfnamefont {R.~C.}\ \bibnamefont {Flagan}},
  \ and\ \bibinfo {author} {\bibfnamefont {K.~J.}\ \bibnamefont {Vahala}},\
  }\href@noop {} {\bibfield  {journal} {\bibinfo  {journal} {Science}\ }\textbf
  {\bibinfo {volume} {317}},\ \bibinfo {pages} {783} (\bibinfo {year}
  {2007})}\BibitemShut {NoStop}%
\bibitem [{\citenamefont {Vollmer}\ \emph {et~al.}(2008)\citenamefont
  {Vollmer}, \citenamefont {Arnold},\ and\ \citenamefont
  {Keng}}]{Vollmer:2008}%
  \BibitemOpen
  \bibfield  {author} {\bibinfo {author} {\bibfnamefont {F.}~\bibnamefont
  {Vollmer}}, \bibinfo {author} {\bibfnamefont {S.}~\bibnamefont {Arnold}}, \
  and\ \bibinfo {author} {\bibfnamefont {D.}~\bibnamefont {Keng}},\ }\href@noop
  {} {\bibfield  {journal} {\bibinfo  {journal} {Proc. Natl. Acad. Sci. USA}\
  }\textbf {\bibinfo {volume} {105}},\ \bibinfo {pages} {20701} (\bibinfo
  {year} {2008})}\BibitemShut {NoStop}%
\bibitem [{\citenamefont {Kippenberg}\ \emph {et~al.}(2011)\citenamefont
  {Kippenberg}, \citenamefont {Holzwarth},\ and\ \citenamefont
  {Diddams}}]{kippenberg2011microresonator}%
  \BibitemOpen
  \bibfield  {author} {\bibinfo {author} {\bibfnamefont {T.}~\bibnamefont
  {Kippenberg}}, \bibinfo {author} {\bibfnamefont {R.}~\bibnamefont
  {Holzwarth}}, \ and\ \bibinfo {author} {\bibfnamefont {S.}~\bibnamefont
  {Diddams}},\ }\href@noop {} {\bibfield  {journal} {\bibinfo  {journal}
  {Science}\ }\textbf {\bibinfo {volume} {332}},\ \bibinfo {pages} {555}
  (\bibinfo {year} {2011})}\BibitemShut {NoStop}%
\bibitem [{\citenamefont {Matsko}\ \emph {et~al.}(2007)\citenamefont {Matsko},
  \citenamefont {Savchenkov}, \citenamefont {Yu},\ and\ \citenamefont
  {Maleki}}]{matsko2007whispering}%
  \BibitemOpen
  \bibfield  {author} {\bibinfo {author} {\bibfnamefont {A.~B.}\ \bibnamefont
  {Matsko}}, \bibinfo {author} {\bibfnamefont {A.~A.}\ \bibnamefont
  {Savchenkov}}, \bibinfo {author} {\bibfnamefont {N.}~\bibnamefont {Yu}}, \
  and\ \bibinfo {author} {\bibfnamefont {L.}~\bibnamefont {Maleki}},\
  }\href@noop {} {\bibfield  {journal} {\bibinfo  {journal} {J. Opt. Soc. Am.
  B}\ }\textbf {\bibinfo {volume} {24}},\ \bibinfo {pages} {1324} (\bibinfo
  {year} {2007})}\BibitemShut {NoStop}%
\bibitem [{\citenamefont {Kessler}\ \emph {et~al.}(2012)\citenamefont
  {Kessler}, \citenamefont {Hagemann}, \citenamefont {Grebing}, \citenamefont
  {Legero}, \citenamefont {Sterr}, \citenamefont {Riehle}, \citenamefont
  {Martin}, \citenamefont {Chen},\ and\ \citenamefont {Ye}}]{kessler2012sub}%
  \BibitemOpen
  \bibfield  {author} {\bibinfo {author} {\bibfnamefont {T.}~\bibnamefont
  {Kessler}}, \bibinfo {author} {\bibfnamefont {C.}~\bibnamefont {Hagemann}},
  \bibinfo {author} {\bibfnamefont {C.}~\bibnamefont {Grebing}}, \bibinfo
  {author} {\bibfnamefont {T.}~\bibnamefont {Legero}}, \bibinfo {author}
  {\bibfnamefont {U.}~\bibnamefont {Sterr}}, \bibinfo {author} {\bibfnamefont
  {F.}~\bibnamefont {Riehle}}, \bibinfo {author} {\bibfnamefont
  {M.}~\bibnamefont {Martin}}, \bibinfo {author} {\bibfnamefont
  {L.}~\bibnamefont {Chen}}, \ and\ \bibinfo {author} {\bibfnamefont
  {J.}~\bibnamefont {Ye}},\ }\href@noop {} {\bibfield  {journal} {\bibinfo
  {journal} {Nat. Photonics}\ }\textbf {\bibinfo {volume} {6}},\ \bibinfo
  {pages} {687} (\bibinfo {year} {2012})}\BibitemShut {NoStop}%
\bibitem [{\citenamefont {Amairi}\ \emph {et~al.}(2012)\citenamefont {Amairi},
  \citenamefont {Legero}, \citenamefont {Kessler}, \citenamefont {Sterr},
  \citenamefont {W{\"u}bbena}, \citenamefont {Mandel},\ and\ \citenamefont
  {Schmidt}}]{amairi2012reducing}%
  \BibitemOpen
  \bibfield  {author} {\bibinfo {author} {\bibfnamefont {S.}~\bibnamefont
  {Amairi}}, \bibinfo {author} {\bibfnamefont {T.}~\bibnamefont {Legero}},
  \bibinfo {author} {\bibfnamefont {T.}~\bibnamefont {Kessler}}, \bibinfo
  {author} {\bibfnamefont {U.}~\bibnamefont {Sterr}}, \bibinfo {author}
  {\bibfnamefont {J.~B.}\ \bibnamefont {W{\"u}bbena}}, \bibinfo {author}
  {\bibfnamefont {O.}~\bibnamefont {Mandel}}, \ and\ \bibinfo {author}
  {\bibfnamefont {P.~O.}\ \bibnamefont {Schmidt}},\ }\href@noop {} {\bibfield
  {journal} {\bibinfo  {journal} {Appl. Phys. B}\ ,\ \bibinfo {pages} {1}}
  (\bibinfo {year} {2012})}\BibitemShut {NoStop}%
\bibitem [{\citenamefont {Numata}\ \emph {et~al.}(2004)\citenamefont {Numata},
  \citenamefont {Kemery},\ and\ \citenamefont {Camp}}]{numata2004thermal}%
  \BibitemOpen
  \bibfield  {author} {\bibinfo {author} {\bibfnamefont {K.}~\bibnamefont
  {Numata}}, \bibinfo {author} {\bibfnamefont {A.}~\bibnamefont {Kemery}}, \
  and\ \bibinfo {author} {\bibfnamefont {J.}~\bibnamefont {Camp}},\ }\href@noop
  {} {\bibfield  {journal} {\bibinfo  {journal} {Phys. Rev. Lett.}\ }\textbf
  {\bibinfo {volume} {93}},\ \bibinfo {pages} {250602} (\bibinfo {year}
  {2004})}\BibitemShut {NoStop}%
\bibitem [{\citenamefont {Webster}\ \emph {et~al.}(2008)\citenamefont
  {Webster}, \citenamefont {Oxborrow}, \citenamefont {Pugla}, \citenamefont
  {Millo},\ and\ \citenamefont {Gill}}]{webster2008thermal}%
  \BibitemOpen
  \bibfield  {author} {\bibinfo {author} {\bibfnamefont {S.~A.}\ \bibnamefont
  {Webster}}, \bibinfo {author} {\bibfnamefont {M.}~\bibnamefont {Oxborrow}},
  \bibinfo {author} {\bibfnamefont {S.}~\bibnamefont {Pugla}}, \bibinfo
  {author} {\bibfnamefont {J.}~\bibnamefont {Millo}}, \ and\ \bibinfo {author}
  {\bibfnamefont {P.}~\bibnamefont {Gill}},\ }\href@noop {} {\bibfield
  {journal} {\bibinfo  {journal} {Phys. Rev. A}\ }\textbf {\bibinfo {volume}
  {77}},\ \bibinfo {pages} {033847} (\bibinfo {year} {2008})}\BibitemShut
  {NoStop}%
\bibitem [{\citenamefont {Millo}\ \emph {et~al.}(2009)\citenamefont {Millo},
  \citenamefont {Magalhaes}, \citenamefont {Mandache}, \citenamefont {Le~Coq},
  \citenamefont {English}, \citenamefont {Westergaard}, \citenamefont
  {Lodewyck}, \citenamefont {Bize}, \citenamefont {Lemonde},\ and\
  \citenamefont {Santarelli}}]{millo2009ultrastable}%
  \BibitemOpen
  \bibfield  {author} {\bibinfo {author} {\bibfnamefont {J.}~\bibnamefont
  {Millo}}, \bibinfo {author} {\bibfnamefont {D.~V.}\ \bibnamefont
  {Magalhaes}}, \bibinfo {author} {\bibfnamefont {C.}~\bibnamefont {Mandache}},
  \bibinfo {author} {\bibfnamefont {Y.}~\bibnamefont {Le~Coq}}, \bibinfo
  {author} {\bibfnamefont {E.~M.~L.}\ \bibnamefont {English}}, \bibinfo
  {author} {\bibfnamefont {P.~G.}\ \bibnamefont {Westergaard}}, \bibinfo
  {author} {\bibfnamefont {J.}~\bibnamefont {Lodewyck}}, \bibinfo {author}
  {\bibfnamefont {S.}~\bibnamefont {Bize}}, \bibinfo {author} {\bibfnamefont
  {P.}~\bibnamefont {Lemonde}}, \ and\ \bibinfo {author} {\bibfnamefont
  {G.}~\bibnamefont {Santarelli}},\ }\href@noop {} {\bibfield  {journal}
  {\bibinfo  {journal} {Phys. Rev. A}\ }\textbf {\bibinfo {volume} {79}},\
  \bibinfo {pages} {053829} (\bibinfo {year} {2009})}\BibitemShut {NoStop}%
\bibitem [{\citenamefont {Strekalov}\ \emph {et~al.}(2011)\citenamefont
  {Strekalov}, \citenamefont {Thompson}, \citenamefont {Baumgartel},
  \citenamefont {Grudinin},\ and\ \citenamefont {Yu}}]{Strekalov:11}%
  \BibitemOpen
  \bibfield  {author} {\bibinfo {author} {\bibfnamefont {D.~V.}\ \bibnamefont
  {Strekalov}}, \bibinfo {author} {\bibfnamefont {R.~J.}\ \bibnamefont
  {Thompson}}, \bibinfo {author} {\bibfnamefont {L.~M.}\ \bibnamefont
  {Baumgartel}}, \bibinfo {author} {\bibfnamefont {I.~S.}\ \bibnamefont
  {Grudinin}}, \ and\ \bibinfo {author} {\bibfnamefont {N.}~\bibnamefont
  {Yu}},\ }\href {\doibase 10.1364/OE.19.014495} {\bibfield  {journal}
  {\bibinfo  {journal} {Opt. Express}\ }\textbf {\bibinfo {volume} {19}},\
  \bibinfo {pages} {14495} (\bibinfo {year} {2011})}\BibitemShut {NoStop}%
\bibitem [{\citenamefont {Fescenko}\ \emph {et~al.}(2012)\citenamefont
  {Fescenko}, \citenamefont {Alnis}, \citenamefont {Schliesser}, \citenamefont
  {Wang}, \citenamefont {Kippenberg},\ and\ \citenamefont
  {H{\"a}nsch}}]{fescenko2012dual}%
  \BibitemOpen
  \bibfield  {author} {\bibinfo {author} {\bibfnamefont {I.}~\bibnamefont
  {Fescenko}}, \bibinfo {author} {\bibfnamefont {J.}~\bibnamefont {Alnis}},
  \bibinfo {author} {\bibfnamefont {A.}~\bibnamefont {Schliesser}}, \bibinfo
  {author} {\bibfnamefont {C.}~\bibnamefont {Wang}}, \bibinfo {author}
  {\bibfnamefont {T.}~\bibnamefont {Kippenberg}}, \ and\ \bibinfo {author}
  {\bibfnamefont {T.}~\bibnamefont {H{\"a}nsch}},\ }\href@noop {} {\bibfield
  {journal} {\bibinfo  {journal} {Opt. Express}\ }\textbf {\bibinfo {volume}
  {20}},\ \bibinfo {pages} {19185} (\bibinfo {year} {2012})}\BibitemShut
  {NoStop}%
\bibitem [{\citenamefont {Baumgartel}\ \emph {et~al.}(2012)\citenamefont
  {Baumgartel}, \citenamefont {Thompson},\ and\ \citenamefont
  {Yu}}]{Baumgartel:12}%
  \BibitemOpen
  \bibfield  {author} {\bibinfo {author} {\bibfnamefont {L.~M.}\ \bibnamefont
  {Baumgartel}}, \bibinfo {author} {\bibfnamefont {R.~J.}\ \bibnamefont
  {Thompson}}, \ and\ \bibinfo {author} {\bibfnamefont {N.}~\bibnamefont
  {Yu}},\ }\href {\doibase 10.1364/OE.20.029798} {\bibfield  {journal}
  {\bibinfo  {journal} {Opt. Express}\ }\textbf {\bibinfo {volume} {20}},\
  \bibinfo {pages} {29798} (\bibinfo {year} {2012})}\BibitemShut {NoStop}%
\bibitem [{\citenamefont {Savchenkov}\ \emph
  {et~al.}(2007{\natexlab{a}})\citenamefont {Savchenkov}, \citenamefont
  {Matsko}, \citenamefont {Ilchenko},\ and\ \citenamefont
  {Maleki}}]{savchenkov2007optical}%
  \BibitemOpen
  \bibfield  {author} {\bibinfo {author} {\bibfnamefont {A.~A.}\ \bibnamefont
  {Savchenkov}}, \bibinfo {author} {\bibfnamefont {A.~B.}\ \bibnamefont
  {Matsko}}, \bibinfo {author} {\bibfnamefont {V.~S.}\ \bibnamefont
  {Ilchenko}}, \ and\ \bibinfo {author} {\bibfnamefont {L.}~\bibnamefont
  {Maleki}},\ }\href@noop {} {\bibfield  {journal} {\bibinfo  {journal} {Opt.
  Express}\ }\textbf {\bibinfo {volume} {15}},\ \bibinfo {pages} {6768}
  (\bibinfo {year} {2007}{\natexlab{a}})}\BibitemShut {NoStop}%
\bibitem [{\citenamefont {Braginsky}\ \emph {et~al.}(1999)\citenamefont
  {Braginsky}, \citenamefont {Gorodetsky},\ and\ \citenamefont
  {Vyatchanin}}]{braginsky1999thermodynamical}%
  \BibitemOpen
  \bibfield  {author} {\bibinfo {author} {\bibfnamefont {V.}~\bibnamefont
  {Braginsky}}, \bibinfo {author} {\bibfnamefont {M.}~\bibnamefont
  {Gorodetsky}}, \ and\ \bibinfo {author} {\bibfnamefont {S.}~\bibnamefont
  {Vyatchanin}},\ }\href@noop {} {\bibfield  {journal} {\bibinfo  {journal}
  {Phys. Lett. A}\ }\textbf {\bibinfo {volume} {264}},\ \bibinfo {pages} {1}
  (\bibinfo {year} {1999})}\BibitemShut {NoStop}%
\bibitem [{\citenamefont {Gorodetsky}\ and\ \citenamefont
  {Grudinin}(2004)}]{gorodetsky2004fundamental}%
  \BibitemOpen
  \bibfield  {author} {\bibinfo {author} {\bibfnamefont {M.~L.}\ \bibnamefont
  {Gorodetsky}}\ and\ \bibinfo {author} {\bibfnamefont {I.~S.}\ \bibnamefont
  {Grudinin}},\ }\href@noop {} {\bibfield  {journal} {\bibinfo  {journal} {J.
  Opt. Soc. Am. B}\ }\textbf {\bibinfo {volume} {21}},\ \bibinfo {pages} {697}
  (\bibinfo {year} {2004})}\BibitemShut {NoStop}%
\bibitem [{\citenamefont {Schliesser}(2009)}]{schliesser}%
  \BibitemOpen
  \bibfield  {author} {\bibinfo {author} {\bibfnamefont {A.}~\bibnamefont
  {Schliesser}},\ }\emph {\bibinfo {title} {Cavity Optomechanics and Optical
  Frequency Comb Generation with Silica Whispering-Gallery-Mode
  Microresonators}},\ \href@noop {} {Ph.D. thesis},\ \bibinfo  {school} {Ludwig
  Maximilian University of Munich} (\bibinfo {year} {2009})\BibitemShut
  {NoStop}%
\bibitem [{\citenamefont {Little}\ \emph {et~al.}(1999)\citenamefont {Little},
  \citenamefont {Laine},\ and\ \citenamefont {Haus}}]{little1999analytic}%
  \BibitemOpen
  \bibfield  {author} {\bibinfo {author} {\bibfnamefont {B.~E.}\ \bibnamefont
  {Little}}, \bibinfo {author} {\bibfnamefont {J.-P.}\ \bibnamefont {Laine}}, \
  and\ \bibinfo {author} {\bibfnamefont {H.~A.}\ \bibnamefont {Haus}},\
  }\href@noop {} {\bibfield  {journal} {\bibinfo  {journal} {J. Lightwave
  Technol.}\ }\textbf {\bibinfo {volume} {17}},\ \bibinfo {pages} {704}
  (\bibinfo {year} {1999})}\BibitemShut {NoStop}%
\bibitem [{\citenamefont {Drever}\ \emph {et~al.}(1983)\citenamefont {Drever},
  \citenamefont {Hall}, \citenamefont {Kowalski}, \citenamefont {Hough},
  \citenamefont {Ford}, \citenamefont {Munley},\ and\ \citenamefont
  {Ward}}]{drever1983laser}%
  \BibitemOpen
  \bibfield  {author} {\bibinfo {author} {\bibfnamefont {R.}~\bibnamefont
  {Drever}}, \bibinfo {author} {\bibfnamefont {J.~L.}\ \bibnamefont {Hall}},
  \bibinfo {author} {\bibfnamefont {F.}~\bibnamefont {Kowalski}}, \bibinfo
  {author} {\bibfnamefont {J.}~\bibnamefont {Hough}}, \bibinfo {author}
  {\bibfnamefont {G.}~\bibnamefont {Ford}}, \bibinfo {author} {\bibfnamefont
  {A.}~\bibnamefont {Munley}}, \ and\ \bibinfo {author} {\bibfnamefont
  {H.}~\bibnamefont {Ward}},\ }\href@noop {} {\bibfield  {journal} {\bibinfo
  {journal} {Appl. Phys. B}\ }\textbf {\bibinfo {volume} {31}},\ \bibinfo
  {pages} {97} (\bibinfo {year} {1983})}\BibitemShut {NoStop}%
\bibitem [{Note1()}]{Note1}%
  \BibitemOpen
  \bibinfo {note} {We note that suppressing mode-averaged temperature
  fluctuations using this technique does not significantly improve the absolute
  frequency stability of the WGM resonator. Eq.~\ref {modefreq} shows that the
  mode frequency depends on the temperature distribution throughout the
  resonator, which is not fixed by mode-averaged temperature control \cite
  {fescenko2012dual,Baumgartel:12}.}\BibitemShut {Stop}%
\bibitem [{\citenamefont {Ghulinyan}\ \emph {et~al.}(2013)\citenamefont
  {Ghulinyan}, \citenamefont {Ramiro-Manzano}, \citenamefont {Prtljaga},
  \citenamefont {Guider}, \citenamefont {Carusotto}, \citenamefont {Pitanti},
  \citenamefont {Pucker},\ and\ \citenamefont
  {Pavesi}}]{Ghulinyan2013oscillatory}%
  \BibitemOpen
  \bibfield  {author} {\bibinfo {author} {\bibfnamefont {M.}~\bibnamefont
  {Ghulinyan}}, \bibinfo {author} {\bibfnamefont {F.}~\bibnamefont
  {Ramiro-Manzano}}, \bibinfo {author} {\bibfnamefont {N.}~\bibnamefont
  {Prtljaga}}, \bibinfo {author} {\bibfnamefont {R.}~\bibnamefont {Guider}},
  \bibinfo {author} {\bibfnamefont {I.}~\bibnamefont {Carusotto}}, \bibinfo
  {author} {\bibfnamefont {A.}~\bibnamefont {Pitanti}}, \bibinfo {author}
  {\bibfnamefont {G.}~\bibnamefont {Pucker}}, \ and\ \bibinfo {author}
  {\bibfnamefont {L.}~\bibnamefont {Pavesi}},\ }\href {\doibase
  10.1103/PhysRevLett.110.163901} {\bibfield  {journal} {\bibinfo  {journal}
  {Phys. Rev. Lett.}\ }\textbf {\bibinfo {volume} {110}},\ \bibinfo {pages}
  {163901} (\bibinfo {year} {2013})}\BibitemShut {NoStop}%
\bibitem [{\citenamefont {Goda}\ \emph {et~al.}(2005)\citenamefont {Goda},
  \citenamefont {McKenzie}, \citenamefont {Mikhailov}, \citenamefont {Lam},
  \citenamefont {McClelland},\ and\ \citenamefont
  {Mavalvala}}]{goda2005photothermal}%
  \BibitemOpen
  \bibfield  {author} {\bibinfo {author} {\bibfnamefont {K.}~\bibnamefont
  {Goda}}, \bibinfo {author} {\bibfnamefont {K.}~\bibnamefont {McKenzie}},
  \bibinfo {author} {\bibfnamefont {E.~E.}\ \bibnamefont {Mikhailov}}, \bibinfo
  {author} {\bibfnamefont {P.~K.}\ \bibnamefont {Lam}}, \bibinfo {author}
  {\bibfnamefont {D.~E.}\ \bibnamefont {McClelland}}, \ and\ \bibinfo {author}
  {\bibfnamefont {N.}~\bibnamefont {Mavalvala}},\ }\href@noop {} {\bibfield
  {journal} {\bibinfo  {journal} {Phys. Rev. A}\ }\textbf {\bibinfo {volume}
  {72}},\ \bibinfo {pages} {043819} (\bibinfo {year} {2005})}\BibitemShut
  {NoStop}%
\bibitem [{\citenamefont {Savchenkov}\ \emph
  {et~al.}(2007{\natexlab{b}})\citenamefont {Savchenkov}, \citenamefont
  {Matsko}, \citenamefont {Ilchenko}, \citenamefont {Yu},\ and\ \citenamefont
  {Maleki}}]{savchenkov2007whispering}%
  \BibitemOpen
  \bibfield  {author} {\bibinfo {author} {\bibfnamefont {A.~A.}\ \bibnamefont
  {Savchenkov}}, \bibinfo {author} {\bibfnamefont {A.~B.}\ \bibnamefont
  {Matsko}}, \bibinfo {author} {\bibfnamefont {V.~S.}\ \bibnamefont
  {Ilchenko}}, \bibinfo {author} {\bibfnamefont {N.}~\bibnamefont {Yu}}, \ and\
  \bibinfo {author} {\bibfnamefont {L.}~\bibnamefont {Maleki}},\ }\href@noop {}
  {\bibfield  {journal} {\bibinfo  {journal} {J. Opt. Soc. Am. B}\ }\textbf
  {\bibinfo {volume} {24}},\ \bibinfo {pages} {2988} (\bibinfo {year}
  {2007}{\natexlab{b}})}\BibitemShut {NoStop}%
\bibitem [{\citenamefont {Feldman}\ \emph {et~al.}(1979)\citenamefont
  {Feldman}, \citenamefont {Horowitz}, \citenamefont {Waxler},\ and\
  \citenamefont {Dodge}}]{feldman1979optical}%
  \BibitemOpen
  \bibfield  {author} {\bibinfo {author} {\bibfnamefont {A.}~\bibnamefont
  {Feldman}}, \bibinfo {author} {\bibfnamefont {D.}~\bibnamefont {Horowitz}},
  \bibinfo {author} {\bibfnamefont {R.~M.}\ \bibnamefont {Waxler}}, \ and\
  \bibinfo {author} {\bibfnamefont {M.~J.}\ \bibnamefont {Dodge}},\ }\href@noop
  {} {\emph {\bibinfo {title} {Optical Materials Characterization, Final
  Technical Report February 1, 1978-September 30, 1978}}},\ \bibinfo {type}
  {Tech. Rep.}\ (\bibinfo  {institution} {DTIC Document},\ \bibinfo {year}
  {1979})\BibitemShut {NoStop}%
\bibitem [{\citenamefont {Allan}(1966)}]{allan1966statistics}%
  \BibitemOpen
  \bibfield  {author} {\bibinfo {author} {\bibfnamefont {D.~W.}\ \bibnamefont
  {Allan}},\ }\href@noop {} {\bibfield  {journal} {\bibinfo  {journal} {Proc.
  IEEE}\ }\textbf {\bibinfo {volume} {54}},\ \bibinfo {pages} {221} (\bibinfo
  {year} {1966})}\BibitemShut {NoStop}%
\bibitem [{\citenamefont {Ghosh}(1998)}]{ghosh1998handbook}%
  \BibitemOpen
  \bibfield  {author} {\bibinfo {author} {\bibfnamefont {G.}~\bibnamefont
  {Ghosh}},\ }\href@noop {} {\emph {\bibinfo {title} {Handbook of Optical
  Constants of Solids: Handbook of Thermo-Optic Coefficients of Optical
  Materials with Applications}}}\ (\bibinfo  {publisher} {Academic Press},\
  \bibinfo {year} {1998})\BibitemShut {NoStop}%
\end{thebibliography}%

\end{document}